# Advances towards a General-Purpose Societal-Scale Human-Collective Problem-Solving Engine

Marko A. Rodriguez
Computer Science Department
University of California, Santa Cruz
Santa Cruz, CA, U.S.A
okram@soe.ucsc.edu

**Abstract** – *Human collective intelligence has proved itself as an important factor in a society's ability to accomplish large-scale behavioral feats. As societies have grown in population-size, individuals have seen a decrease in their ability to actively participate in the problem-solving processes of the group. Representative decision-making structures have been used as a modern solution to society's inadequate information-processing infrastructure. With computer and network technologies being further embedded within the fabric of society, the implementation of a general-purpose societal-scale human-collective problem-solving engine is envisioned as a means of furthering the collective-intelligence potential of society. This paper provides both a novel framework for creating collective intelligence systems and a method for implementing a representative and expertise system based on social-network theory.*

**Keywords:** societal-scale decision-support system, e-government, social-network, collective intelligence.

## 1 Introduction

The area of human-collective problem-solving, from the vantage point of computer science, has been studied primarily in the research agenda of group decision-support systems. Work previous to this moment has focused primarily on computer support systems for small groups in a particular problem-domain [2]. It is only recently that research efforts have begun to move away from small-group decision-support systems to societal-scale decision-support systems [12]. In these efforts, large-scale decision-support system development has come to realize necessary research revolving around fluctuating group participation [9], problem modeling [3], decision-making for natural-language problems [12], and general computer-mediated collaboration techniques [8]. The following work provides a conceptual framework for designing societal-scale decision and policy making systems capable of taping the collective intelligence potential of a large group.

## 2 Foundations for Collective Problem-Solving

This section provides a theoretical framework for designing human collective-intelligence systems.

### 2.1 The Environment and the Collective Realization

The 'objective' *environment* shared by a collective of individuals has no inherent property that supports the realization of a 'problem' separate from the internal workings of the group cognition. Therefore, to any subset of the collective, a *problem* is a perceived portion of reality that under the proper transformation could potentially provide a greater utility than presently realized [3, 6]. A collective shares a common understanding of the external environment in so much that the members share the same perceptual mechanisms (physical perceptors as well as internal models such as language and culture) and in so doing are able to come to a realization of a collective-subjective environment. Without a shared environmental realization, the notion of a problem would only be an internal concept represented within the modeling framework of the individual's cognitive faculties. If a problem is to be solved in a collective fashion, there must exist a model of that problem, within a shared medium of the group, by which that problem can be communicated amongst members of the group [4, 7]. Therefore, a *problem-model* is a formal representation, within a shared-medium, of a low-utility aspect of the collective's environment. Individuals then review the problem-model in order to formalize a *solution-model* that satisfies the problem-model's constraints—irrespective of any rippling deleterious effects it may incur on other aspects of the environment. Finally, the *solution* will be defined as the perceptible environmental outcome of the implementation of the formalized solution-model. This lays the foundation for a canonical understanding of how an environment is perceived to have problems that need to be solved via some problem-solving mechanism.

## 2.2 Problem/Solution-Space as the Collective Workspace

As individuals interact with their environment, they go through the process of problem-modeling. The collection of all problem-models within the group can be called the group's *problem-space*. It is within this space that these problem-models are transformed into their respective solution-models to be stored within the group's *solution-space*. The formalized solution-models, within the solution-space, are then implemented as the collectively agreed upon course of action. The problem-space and solution-space can be seen as the collective-workspace by which these humans represent and share their ideas about the world [4, 7]. It is through this shared medium that the cognitive resources of the collective can synergistically coexist to formulate solutions that are greater than what could possibly be achieved by any individual working alone. All italicized concepts mentioned are represented below (*Figure 1*).

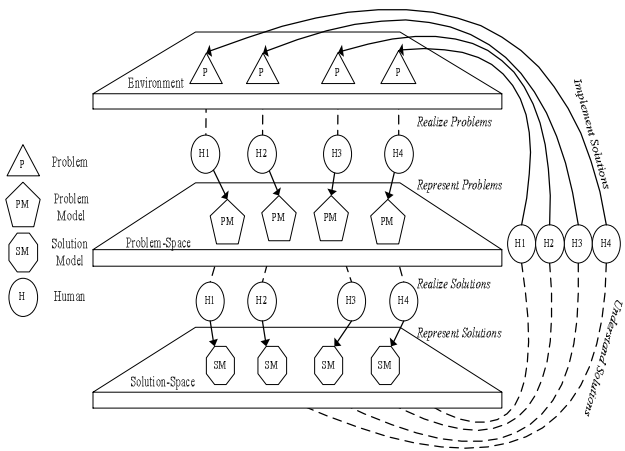

Figure 1: A Canonical Model of a Collective Problem-Solving Process

It is important to note that in the collective problem-solving effort outlined in *Figure 1*, there exists no collaboration among individuals. This simplistic model does not take advantage of the collective's ability to synergistically provide higher-quality solutions, but instead takes advantage of the group synergistic ability to generate a higher-throughput of solutions since environmental problems are solved in a parallel effort. Representing collaborative problem-solving would only complicate a preliminary grasp of the aforementioned concepts.

## 2.3 Problem-Modeling and Decision-Making

In the canonical model presented in *Figure 1*, each individual within the group is capable of formulating a problem-model without collaboration with other group members. In a real-world scenario, this becomes increasingly difficult as the complexity of the problems begin to grow larger than what an individual's mental framework can model. In such cases, problem-modeling may only require one individual to make an initial shallow realization of the problem, but through a more in-depth analysis, spanning various domains, a collaborative effort of domain specialists would be required to create an accurate collective model of the problem. The act of formulating a problem will be called *problem-modeling*. Problem-modeling is an important step in problem-solving since a good representation can make a hard problem trivial, and a bad representation can make a trivial problem hard [3, 6]. For this reason, a selection process must occur in which problem-models are judged, by the group, according to their potential at providing a low-energy effort in the solution-modeling process. The *decision-making process* is a selection process since it doesn't generate new problem-models but instead selects high-quality models. As demonstrated in *Figure 2*, the problem-modeling stage generally sees a growth in the number of problem-models and through the decision-making stage this model pool is pruned down to the most optimal model as seen by the group. Together problem-modeling and decision-making is the *problem-generation process*.

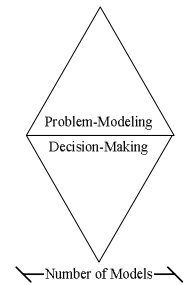

Figure 2: Size of Problem-Model Pool over the Problem-Generation Process

## 2.4 Solution-Modeling and Decision-Making

Once a problem has been modeled within the problem-space, the collective can begin to transform that problem-model into a solution-model ready for implementation in the environment. In order to take advantage of a collaborative effort, two distinct stages must first be traversed. *Solution-modeling* is the process by which individuals review the problem-model, understand the issue at hand, and via the use of their internal cognitive machinery, generate a collection of solution-models that may be of potential use. Hybrid solutions may then be formalized via the act of collaboration. At the end of the solution-modeling process the solution-model pool, like the problem-model pool, must be trimmed via a selection process. In order to yield a single solution-model for the problem, the group must enter a decision-making process in which the group collectively decides, via some *voting algorithm*, which solution-model should be implemented. This two-stage model is coined here as the *solution-generation process*.

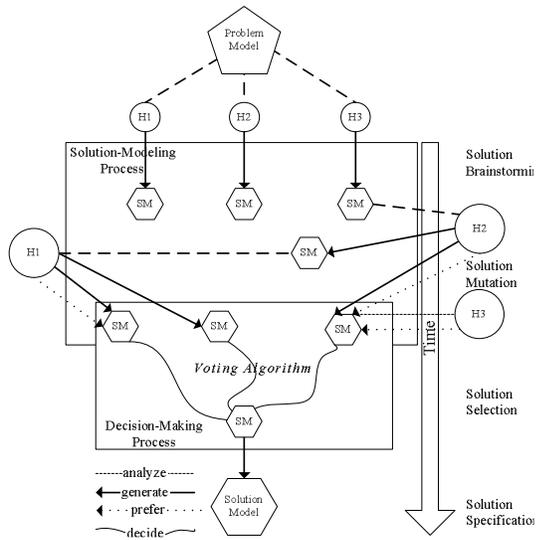

Figure 3: Solution-Generation Process Model

The above example demonstrates how a problem-model is transformed into a solution-model via the solution-generation process. It is important to note that this same principle applies to the way in which a perceived environmental problem is transformed into a problem-model. In *Figure 3*, there are three human participants actively involved in solution-generation. At first, each individual reviews the problem-model and provides their own unique solution-model based solely on their specific internal realizations. Once these three solutions-models have been represented in the solution-space, the individuals are able to analyze each other's models such that they may generate new and potentially higher quality solution-models. After a certain limit (context dependant), the group enters the decision-making process in which the models in the solution pool are voted on. A voting algorithm (context dependant) is able to aggregate the preferences of the individuals to yield a model that is reflective of the group's perspective.

## 2.5 Voting Algorithms and the Context

The areas of social choice theory and machine learning are full of algorithms by which a collective can aggregate the individual perspectives of its members in order to yield a collective perspective [1]. Different algorithms are more appropriate than others depending on the context for which they are needed. For example, where a numeric solution is required, simply averaging the views of all the individuals may be sufficient. For more linguistic models, where numeric averaging isn't possible, a Borda-Count or majority vote over various annotations may be the appropriate method. It is up to the problem-modelers and solution-modelers to decide the appropriate voting algorithm used to filter the model pools.

## 2.6 Problem-Solving as the Unification of all Processes

*Problem-solving* is the unification of all the modeling and decision-making processes described above. This is the general idea that a collective-subjective environmental problem is modeled in a collaborative effort and then solved via a similar mechanism. Problem-solving also encompasses the act of implementing the collective's solution and thus completing the loop—bringing the group closer to equilibrium with its perceived environment.

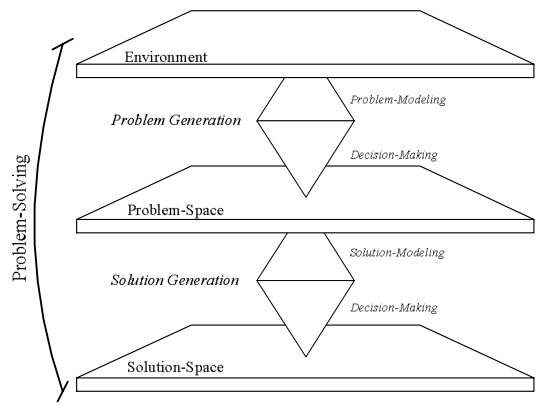

Figure 4: Problem-Solving as the Unification of all Processes

## 3 Group vs. Societal Decision-Support Systems

There has been much literature published in the domain of group decision-support systems. This section provides arguments for why typical group decision-support research differs from societal-scale decision-support research.

## 3.1 Heterogeneity of Individuals and Problems

Group decision-support systems are usually built for a particular problem-solving domain in which the individuals participating in the group decision process are of similar background in terms of the type and level of expertise that they share [12]. More advanced, cross-domain systems do exist to support complex problem-solving, but still at this level, the type of expertise is explicit in the system design. The reason for this is that most decision-support systems rely heavily on human interfaces tailored to a particular domain [8]. A general-purpose system wouldn't have this luxury save through domain-specific extensions to a basic implementation. Also, a general-purpose system hopes to tap collective intelligence on a societal-scale. In doing so, the problems realized by the group and the expertise of the group are extremely diverse in both topic and depth. Therefore, unlike group decision-support system research, a societal-scale decision-support system would need formalized requirements regarding the ad hoc categorization of the expertise of its individuals and their problems as they are realized by the collective. Section five provides a means by which a group can organize the collective-workspace into domain specific categories for both problem organization and representation amongst peers.

## 3.2 Fluctuating Participation and the Collective Perspective

The *collective perspective*, as defined by this work, is the realized course of action by the group in the case that every member participated in the problem-solving process. As members abstain from participation, the ability for the group to maintain that perspective defines how well the collective's self-model is able to handle fluctuating participation levels. With division of labor and parallel decision-making, dealing with fluctuating participation levels is of utmost importance. The collective perspective is made more explicit in opinion-based decision-making such as is seen in value-based political polling. When all individuals actively vote on their opinion, the final decision, provided by the voting algorithm determines the collective's will. If the group maintains a model for holographically representing itself, then as members refrain from voting the same opinion poll would yield the same outcome as the full participation vote. In typical group decision-support system research, the collective perspective is upheld via a requirement of the participation of all members in all decision processes. In these small-scale systems, fluctuating participation can have dramatic effects on the solution quality of the group [11]. Full group participation in small groups is realized via asynchronous support mechanisms. In such scenarios, decision-making usually happens on a time limit in which group members make their choice within a time window, thus ensuring a higher potential for full participation while not requiring simultaneous participation [2]. In a societal-scale system, the ability for the full group to participate in all decision processes is impractical and potentially impossible. In light of this, the collective perspective must be represented without explicit reference to the aggregation of each member's decision. The next section provides a social-network topology and power dissemination algorithm capable of holographically representing the collective perspective over any subset of the active participants in a decision-making process. Thus reducing the error incurred by lossy participation and/or low resolution representative systems.

## 4 Decision-Making Social-Networks

This section presents two distinct simulations of representative social-networks capable of ensuring that the collective's perspective is maintained as member participation is reduced. These social-network algorithms have three important properties. First is the connectivity of the network (K). Human nodes within the network are allowed to select K representatives of their viewpoint. Secondly, node activity (A) determines whether or not a human node is an active participant in the current decision-making process. The connectivity and activity values determine the flow depth (D) of power from non-active members to active participatory members. The accuracy of the network is determined by how well the collective is able to holographically represent itself within the actively participating subset of the group.

These social-networks are presented within the framework of representative decision-making but also work as a means of creating a bottom-up peer-reflective model of the expertise within a group—where edges in the graph represent one's confidence in a peer's ability within a particular problem-domain. Expertise networks support the use of collective-intelligence beyond strict opinion representation. Furthermore, edges can be labeled with problem-domain names such that expertise can be further categorized. Work in expertise modeling and domain modeling on these networks can be found in [10].

### 4.1 Representative Social-Network Simulation

A human is represented as a node within a social-network. Each human in the group is provided with a random opinion between 0.0 and 1.0. A human with a star in their node represents that they are an active participant in a decision-making process of that domain, while non-participants have no star. The first model (*Figure 5*) utilizes a social-network with a single degree of connectedness (K=1) and a power dissemination constraint of depth one (D=1). Power propagates from non-participating individuals to participating individuals over the network edges. Model one requires non-participants to select a single active representative that is closest in opinion—ensuring a power distribution of at most depth one. The value on the outgoing edge represents the percentage of power going to that representative. For this model 100% of the power leaving the non-active node will reach the active representative node. Individuals with power in these systems can be seen has having influence in two areas. First, it can provide weight to an individual's vote strength in decision-making and second, it can allow an individual's models more visibility over less powerful individuals within the collective workspace.

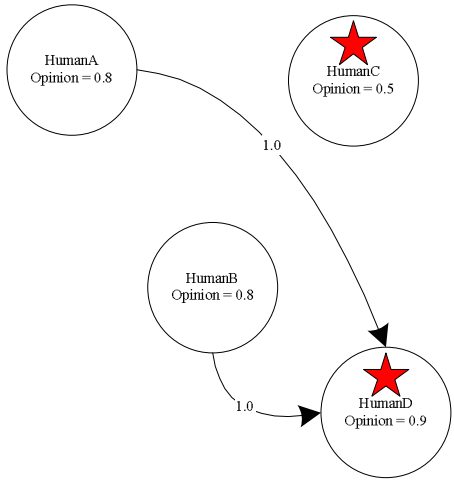

Figure 5: D=1 and K=1 Representative Social-Network

The second model (*Figure 6*) allows an individual to select multiple representatives of their opinion regardless of that representative's active participation—decision power is able to traverse the graph until it reaches an actively participating individual. Power dissemination is bifurcated when an individual has more than one representative. The power bifurcation value is determined by the function below (1). Once the representative edges are determined, the values are normalized to 1.0. The benefit of this model is that member participation can ebb and flow without requiring a rewiring of the connection schema since representatives can be either active or non-active participants.

$$EdgeValue(n_i, n_j) = 1 - |opinion_i - opinion_j| \quad (1)$$

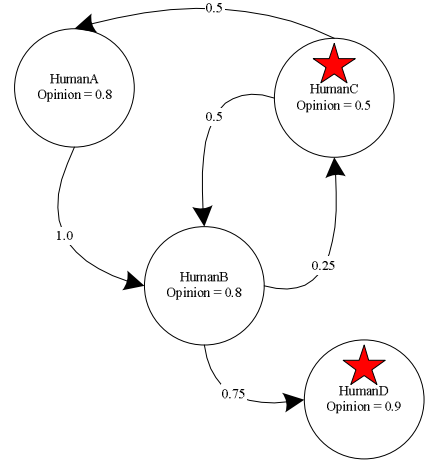

Figure 6: D=X and K=Y Representative Social Network

$$NetworkDecision = \frac{1}{|N|} \sum_{i=1}^{|A|} (power_i * opinion_i) \quad (2)$$

For both social-network decision models, the decision of the collective is determined by averaging the weighted opinions of the active representative decision-makers (2). For comparison, the control simulation is the case in which the collective perspective is perfectly modeled. This is accomplished by averaging the opinions of every member of the group regardless of their active participation (3).

$$PerfectDecision = \frac{1}{|N|} \sum_{i=1}^{|N|} (opinion_i) \quad (3)$$

$$DecisionError = |NetworkDecision - PerfectDecision| \quad (4)$$

The error of each network model's decision is determined by taking the absolute value of the difference between the network model's outcome decision and the perfect decision (4). These simulations were run on a population of 1,000 members with randomly generated opinions. A parameter sweep over the size of the actively participating population was used to test the effects of abstained participation. Various multi-degree networks were tested, from fully connected networks to one-degree networks and the results demonstrated that three-degree networks of variable depth to be the most effective network topology for representative decision-making. They were best able to dampen the error impinged by non-participation (*Figure 7*). Interestingly enough, K=0 networks, where representation is directly related to participation, performs equal to D=1 K=1 networks [10].

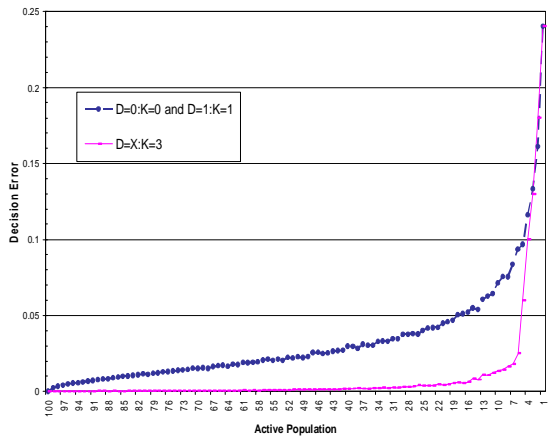

Figure 7: Network Model Decision Errors over Active Participant Percentage

## 5 Conclusions

The ideas presented in the work provide a theoretical model for designing problem-solving systems for societal-scale collectives. Various social-network topologies were discussed as a means of providing a group a representation of its opinion such that any active subset of the population is a model of the perspective of the whole. A modified use of these networks can also be used as a means of creating a bottom-up peer-reflective representation of the relative expertise within a group. The unification of the collective-workspace and the social-network architectures defines the preliminary components needed to develop a human-collective societal-scale problem-solving engine (*Figure 11*).

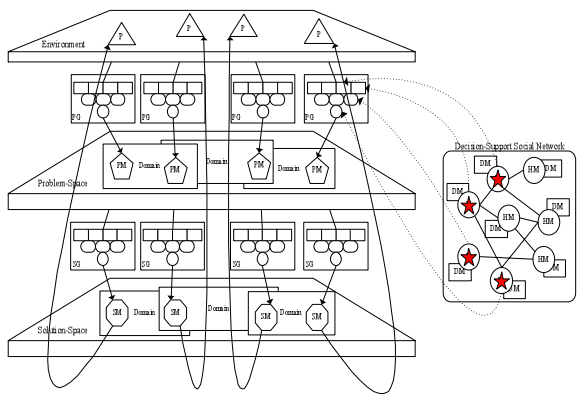

Figure 11: Unification of the Social-Network Model and the Collective-Workspace

Further research and development in this area will lend towards new means by which organizations and political institutions can utilize the collective intelligence potential of a group as a means of providing high-quality solutions to any representable problem facing the group.